\documentclass{elsart}
\usepackage{epsfig,eqnfix}

\newsavebox{\LSIM}
\sbox{\LSIM}{\raisebox{-1ex}{$\ \stackrel{\textstyle<}{\sim}\ $}}

\newsavebox{\GSIM}
\sbox{\GSIM}{\raisebox{-1ex}{$\ \stackrel{\textstyle>}{\sim}\ $}}

\newcommand{\be} {\begin{equation}}
\newcommand{\ee} {\end{equation}}
\newcommand{\bdm} {\begin{displaymath}}
\newcommand{\edm} {\end{displaymath}}
\newcommand{\bc} {\begin{center}}
\newcommand{\ec} {\end{center}}
\newcommand{\beqa} {\begin{eqnarray}}
\newcommand{\eeqa} {\end{eqnarray}}

\newcommand{\ra} {\rightarrow}

\newcommand{\bear}{\begin{eqnarray}}
\newcommand{\ear}{\end{eqnarray}}
\newcommand{\bea}{\begin{eqnarray*}}
\newcommand{\ea}{\end{eqnarray*}}

\newcommand{\slp}{\raise.15ex\hbox{$/$}\kern-.57em\hbox{$\partial$}}
\newcommand{\slG}{\raise.15ex\hbox{$/$}\kern-.57em\hbox{$G$}}
\newcommand{\slA}{\raise.15ex\hbox{$/$}\kern-.57em\hbox{$A$}}
\newcommand{\grgl}{\:\hbox to -0.2pt{\lower2.5pt\hbox{$\sim$}\hss}
{\raise3pt\hbox{$>$}}\:}
\newcommand{\klgl}{\:\hbox to -0.2pt{\lower2.5pt\hbox{$\sim$}\hss}
{\raise3pt\hbox{$<$}}\:}

\newcommand{\befi}[1]{\begin{figure}[ht] \leavevmode \centering \epsffile{#1.eps}}

\newcommand{\beq}{\begin{equation}}
\newcommand{\enq}{\end{equation}}
\newcommand{\beqast}{\begin{eqnarray*}}
\newcommand{\enqa}{\end{eqnarray}}
\newcommand{\enqast}{\end{eqnarray*}}

\def\ct{\cite}
\def\pmb#1{\setbox0=\hbox{#1}% \kern-.025em\copy0\kern-\wd0
\kern.05em\copy0\kern-\wd0 \kern-.025em\raise.0433em\box0 }

\def\ell{l}

\def\P{I\!\!P}

\def\half{{\textstyle{1\over 2}}}

\def\P{I\!\!P}

\begin{document}
\bibliographystyle{plainer}
\begin{frontmatter}
\title{Pomerons}
\author{P V Landshoff}\address{DAMTP, University of Cambridge}
\ead{pvl@damtp.cam.ac.uk}
\begin{abstract}
New small-$x$ data indicate the existence of two pomerons: the usual soft
pomeron with intercept close to 1.08, and a hard pomeron with intercept
about 1.4
\end{abstract}
\maketitle
\end{frontmatter}
\section{The soft pomeron}
\begin{figure}[t]
\begin{center}
\epsfxsize=0.6\textwidth\epsfbox[55 550 366 770]{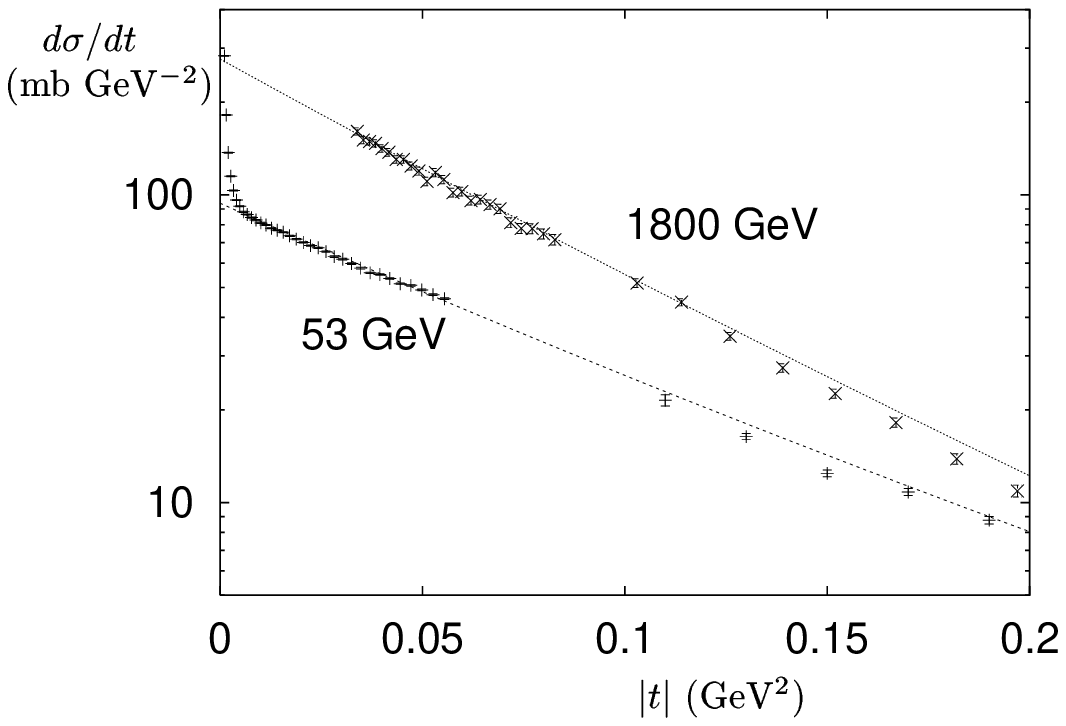}
\end{center}
\caption{$pp$ elastic scattering\cite{Nag79} at $\sqrt s=53$ GeV
and\ct{Amo90} $~\bar pp$ at $\sqrt{s}=1800$ GeV. The curves verify
the value of the soft-pomeron slope to be $\alpha_1'=0.25\hbox{ GeV}^{-2}$.}
\vskip 4truemm
\label{ELASTIC}
\end{figure}

The existence of the soft pomeron is now well-established, even if the
precise value of its intercept $1+\epsilon_1$ is not agreed. Donnachie
and I believe\ct{DL92} that $\epsilon_1$ is close to 0.08, while 
others\ct{CDF94,CEKLT98,PDG00} believe it is close to 0.1. It seems
that, to a good approximation, the trajectory is linear:
\be
\alpha_1(t)=1+\epsilon_1+\alpha'_1t
\label{softpom}
\ee
The value of the slope has been known for a over a quarter of a 
century\ct{JL74} to be 
\be
\alpha_1'=0.25\hbox{ GeV}^{-2}
\label{pomslope}
\ee
This value has since been confirmed by comparison between elastic-scattering
data from the ISR and the Tevatron: see figure \ref{ELASTIC}. The
value (\ref{pomslope}) correctly predicted the observed shrinkage of
the forward peak.
\begin{figure}[t]
\vskip 9truemm
\begin{center}
\epsfxsize=0.5\textwidth
\epsfbox[230 190 430 490]{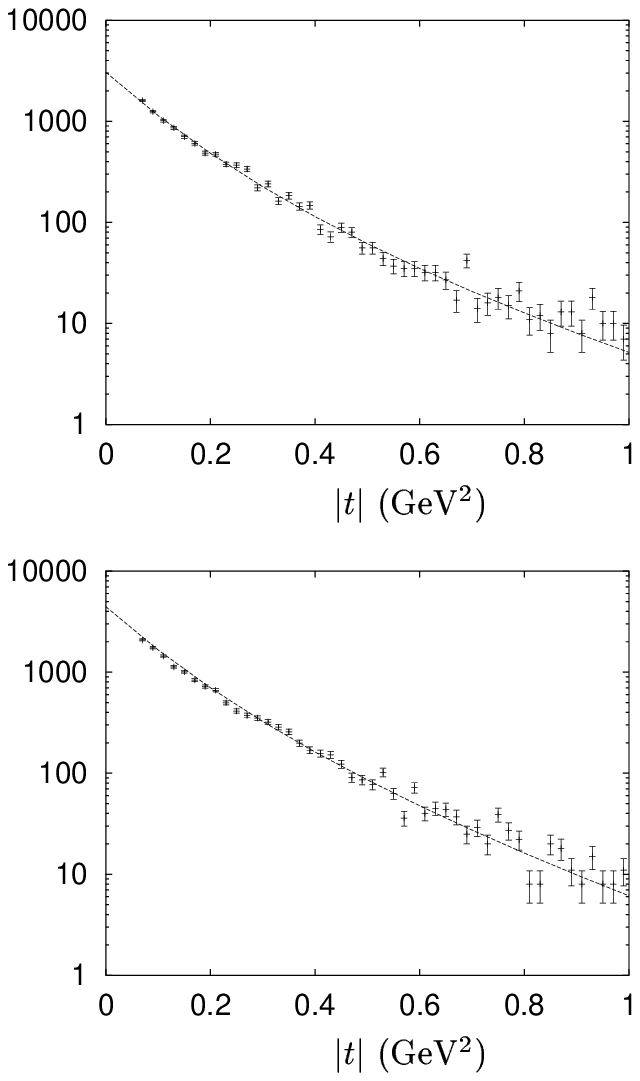}
\end{center}
\vskip -3truemm
\caption{Data\cite{Ast82}
for $\gamma p \ra \rho p$ at $\sqrt s = 6.86$ GeV (upper figure)
and 10.8 GeV (lower figure)
with Regge curves. The data are events per 0.02 GeV$^{-2}$.}
\label{RHODSIGMA1}
\vskip 5truemm
\begin{center}
\epsfxsize=0.55\textwidth\epsfbox[45 564 340 770]{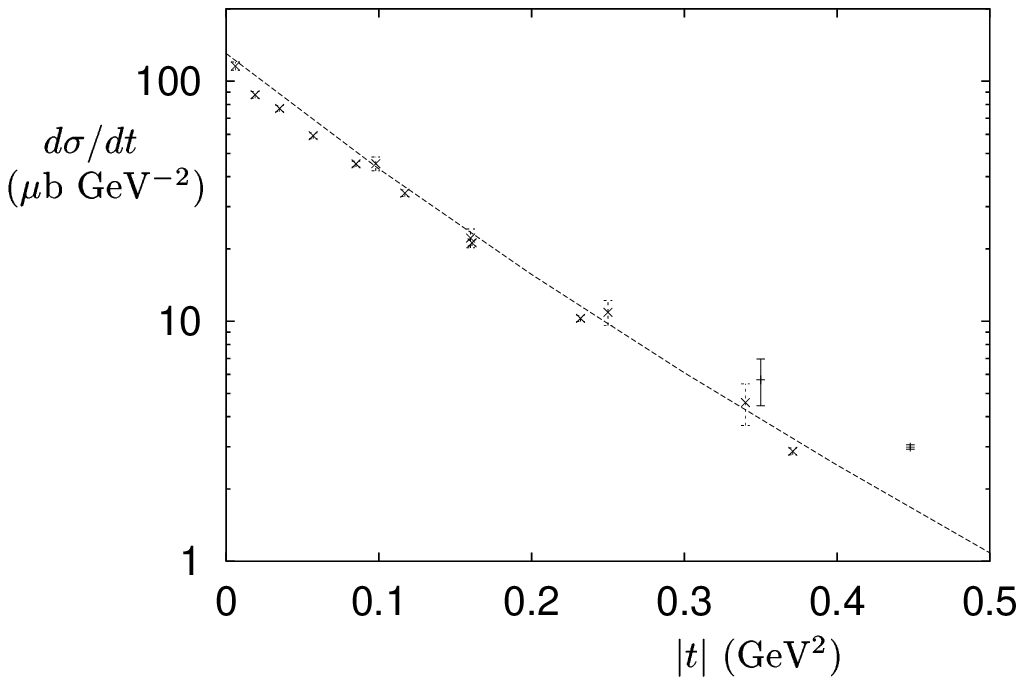}
\end{center}
\caption{ZEUS data\cite{ZEUS98}
for $\gamma p \ra \rho p$ with the Regge-model
prediction. The lower-$t$ data are at $\sqrt{s} = 71.7$ GeV and the higher-$t$
data at \hbox{94 GeV.}}
\vskip 4truemm
\label{ZEUSRHO}
\end{figure}

Another verification comes from the change in the differential cross
section for the process $\gamma p\to\rho p$ with increasing energy.
It is essential to include not just soft-pomeron exchange, but also
$f_2, a_2$ exchange, for which the trajectory is
\be
\alpha_R(t)=1+\epsilon_R+\alpha'_Rt
\label{reggeon}
\ee
with
\be
\epsilon_R\approx -0.5~~~~~~~~~~~\alpha'_R\approx 0.9 \hbox{ Gev}^{-2}
\ee
A fit to the low-energy data is shown in figure \ref{RHODSIGMA1}.
The same fit continues to perform well when extrapolated to high energy:
figure \ref{ZEUSRHO}.

\section{Two pomerons}

While the soft pomeron accounts well for the rise with energy of 
cross sections for soft processes,  data for hard and semi-hard processes
reveal an additional contribution. The data are described very
well\cite{DL98,DL01} by the assumption
that this contribution is a second pomeron. If its trajectory is linear:
\be
\alpha_0(t)=1+\epsilon_0+\alpha'_0t
\label{hardpom1}
\ee
then we find that experiment requires
\be
\epsilon_0\approx 0.4~~~~~~~~~~~\alpha'_0\approx 0.1 \hbox{ Gev}^{-2}
\label{hardpom2}
\ee

\begin{figure}[t]
\begin{center}
\centerline{\hskip -2truemm\epsfxsize=0.484\textwidth\epsfbox[80 590 300 770]{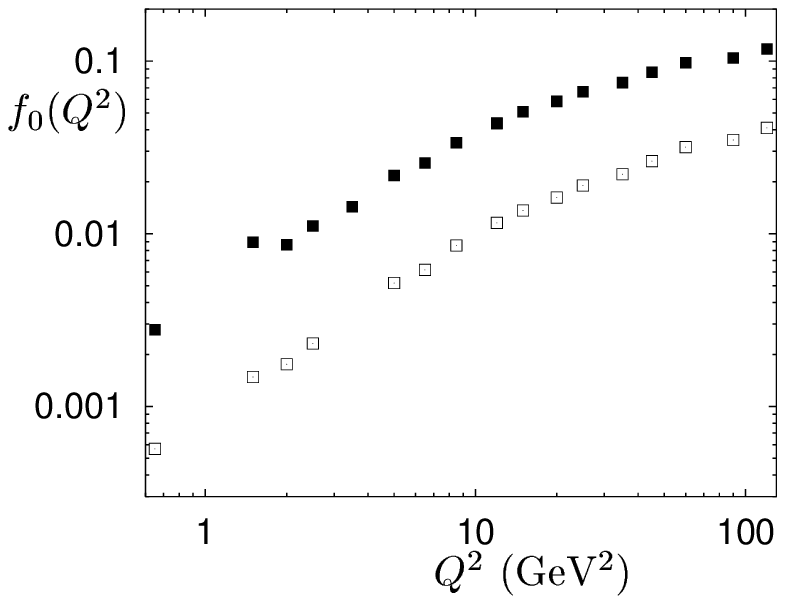}
\hfill
\epsfxsize=0.484\textwidth\epsfbox[80 590 300 770]{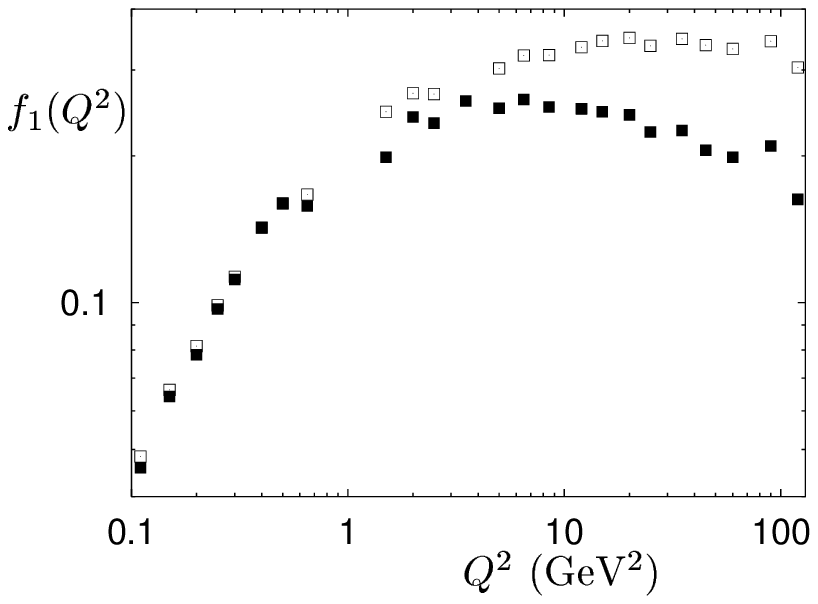}}
\end{center}
\vskip -8truemm
\caption{Fits to the coefficient functions $f_0(Q^2)$ and $f_1(Q^2)$
of (\ref{smallx}) extracted from H1 and ZEUS data\ct{ZEUS00c,H101,H101a}.
The black points
are for $\epsilon_0=0.36$ and the white points are for $\epsilon_0=0.5$.}
\label{COEFF}
\end{figure}

Consider first the proton structure function $F_2(x,Q^2)$. The pomeron exchanges
contribute at small $x$
\be
F_2(x,Q^2\sim\sum_{i=0,1}f_i(Q^2)x^{-\epsilon_i}
\label{smallx}
\ee
Figure \ref{COEFF} shows the coefficient functions $f_i(Q^2)$
extracted from the data\ct{ZEUS00c,H101,H101a}
 for $x<0.02$ at each available value of $Q^2$,
for two choices of $\epsilon_0$. We see that if we suppose that the soft-pomeron
coefficient function $f_1(Q^2)$ becomes constant at large $Q^2$ a larger
value of $\epsilon_0$ is favoured than if we suppose that it decreases
with increasing $Q^2$. We find\ct{DL01} that either hypothesis gives
a good fit to the data. In either case, the hard-pomeron coefficient function
increases at large $Q^2$; the best fit we have found corresponds to a power
behaviour
\be
f_1(Q^2)\sim Q^{\epsilon_0}
\label{power}
\ee
at large $Q^2$. 

\begin{figure}[t]
\vskip -10truemm
\begin{center}
\epsfysize=0.7\textheight\epsfbox[100 300 410 760]{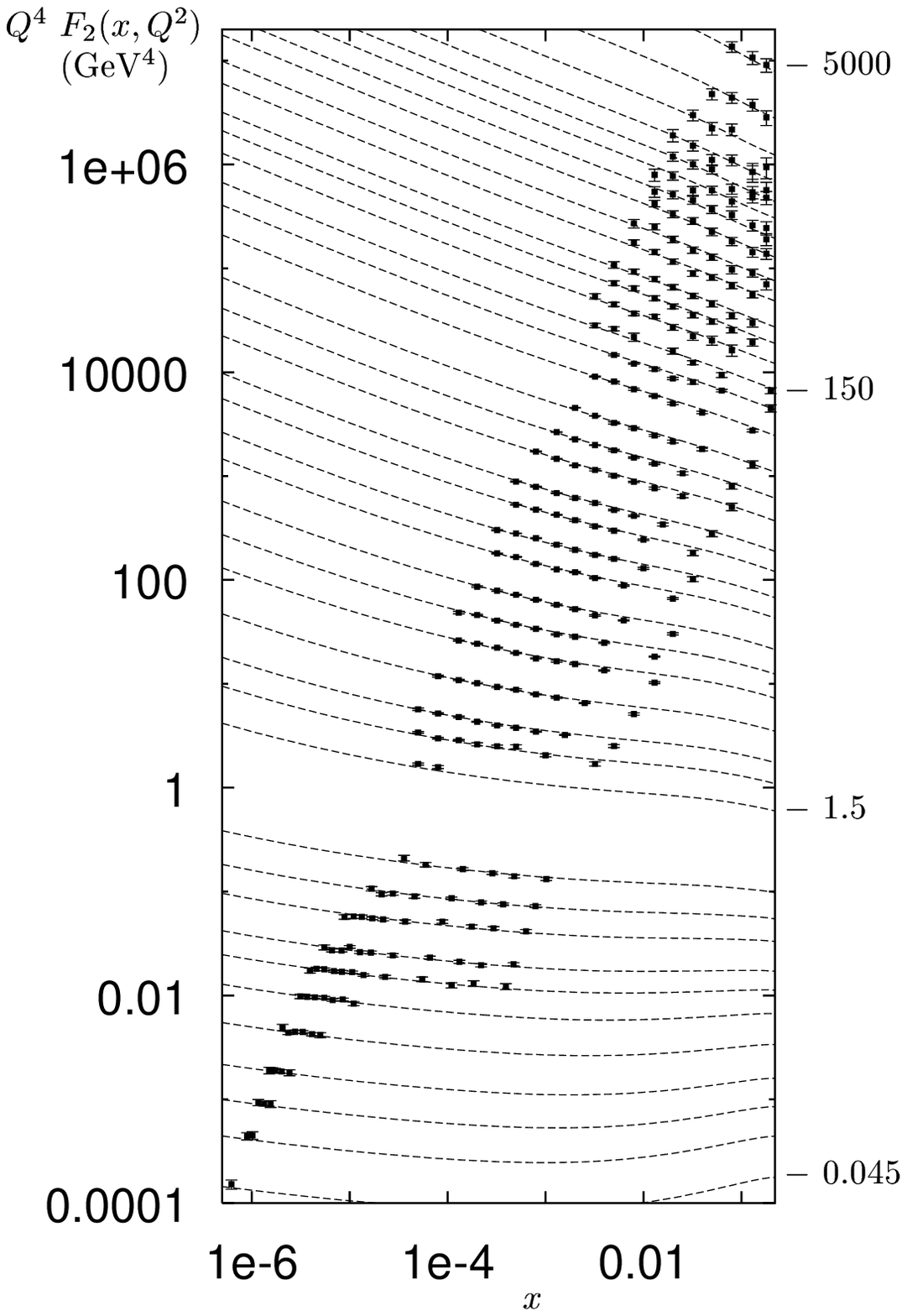}
\end{center}
\vskip -1truemm
\caption{Regge fit to data for $F_2(x,Q^2)$ for $Q^2$ between 0.045
and 5000 GeV$^2$. The parameters were fixed using only data for
$x<0.001$.}
\label{F2}
\end{figure}

The data are now so very accurate that, even at very small $x$, the
simple-power fit~(\ref{smallx}) needs modifying, to take account of the fact
that $F_2(x,Q^2)\to 0$ as $x\to 1$. The simplest assumption, consistent
with the dimensional-counting rules, is to multiply each term
by $(1-x)^7$. This is surely over-simple, but should be better than
not including any such factor. We fit to data for $x<0.001$ and then 
these factors
have a small effect, but it is not completely negligible. The fit is
shown in figure~\ref{F2} and has essentially just 4 free parameters,
including $\epsilon_0$. Although we used only data for $x<0.001$ to
make the fit, we see that it is quite good up to larger $x$, and all the
way from $Q^2=0.045$ to 5000 GeV$^2$. 

\begin{figure}[t]
\vskip -1truemm
\begin{center}
\epsfxsize=0.4\textwidth\epsfbox[85 430 285 760]{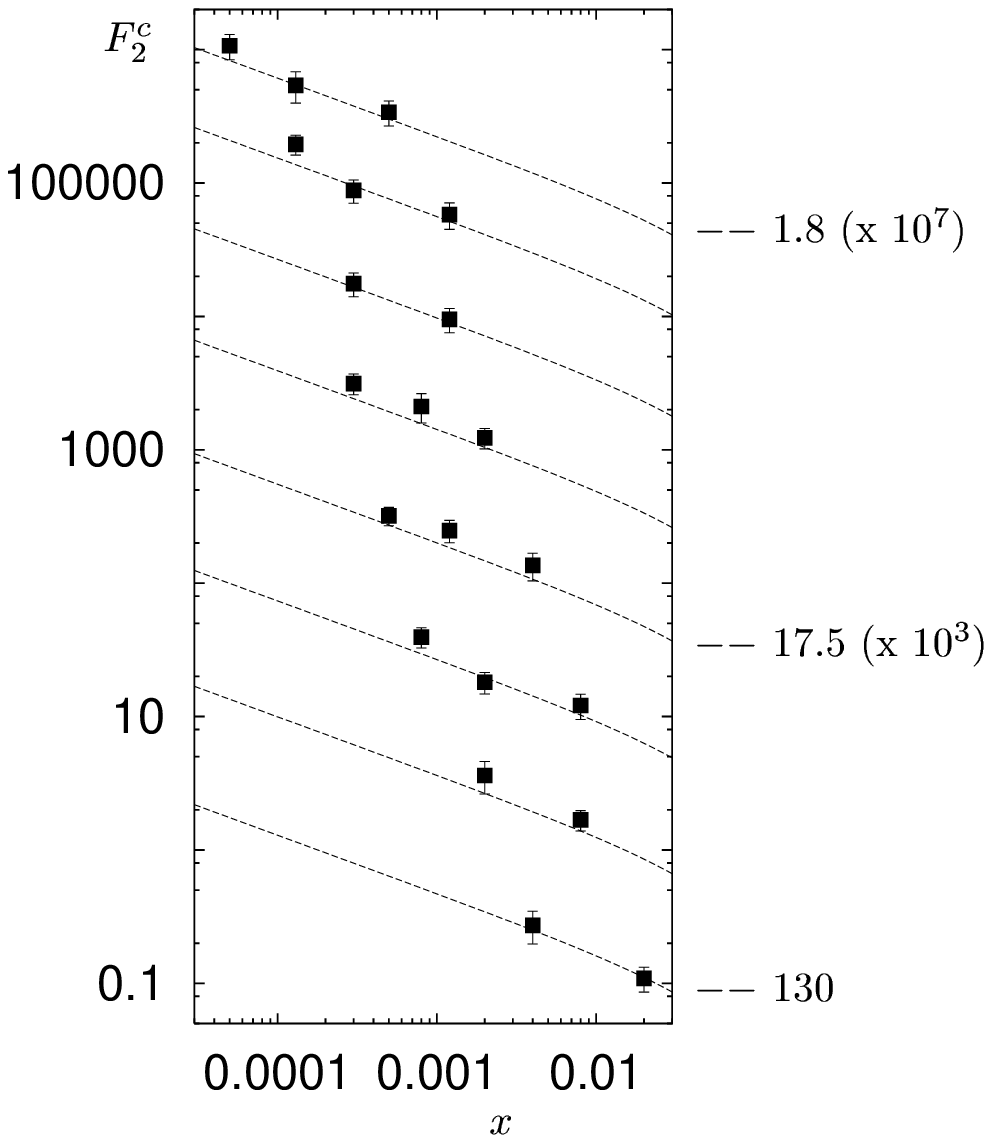}
\end{center}
\vskip -1truemm
\caption{Data\cite{Z00} for $F_2^c(x,Q^2)$ from $Q^2=1.8$ to $130$ GeV$^2$.
The curves are the hard-pomeron component of the fit to $F_2(x,Q^2)$
shown in figure \ref{F2}, normalised such that the hard pomeron is
flavour-blind.}
\label{CHARM}
\end{figure}

Somewhat remarkably, the data\ct{Z00} for the charm component $F_2^c(x,Q^2)$
of $F_2(x,Q^2)$ have the striking
property
that over a wide range of $Q^2$ they behave as a fixed power
of $x$. They are described very well by a hard-pomeron term alone.
Further, the hard pomeron seems to be flavour-blind. Figure~\ref{CHARM}
shows ZEUS data together with ${2\over 5}$ the hard-pomeron
contribution to $F_2(x,Q^2)$. The factor ${2\over 5}$ is just
$e_c^2/(e_u^2+e_d^2+e_s^2+e_c^2)$. There are considerable uncertainties
about these data, because they are obtained using a very large extrapolation
in $p_T$, but on the face of it they are very compelling evidence for the
hard-pomeron concept.

\section{Real photons}

\begin{figure}[t]
\vskip 8truemm
\begin{center}
\epsfxsize=0.4\textwidth\epsfbox[85 580 300 760]{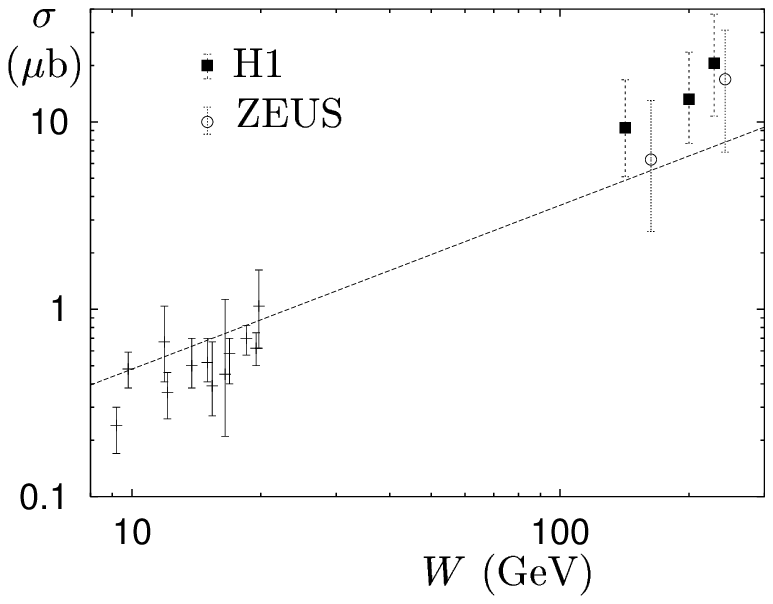}
\end{center}
\vskip -1truemm
\caption{Extrapolation to $Q^2=0$ of the 
hard-pomeron contribution to $F_2^c(x,Q^2)$}

\label{CHARM_PHOTO}
\end{figure}
\begin{figure}[t]
\vskip 3truemm
\begin{center}
\epsfxsize=0.4\textwidth\epsfbox[85 420 280 760]{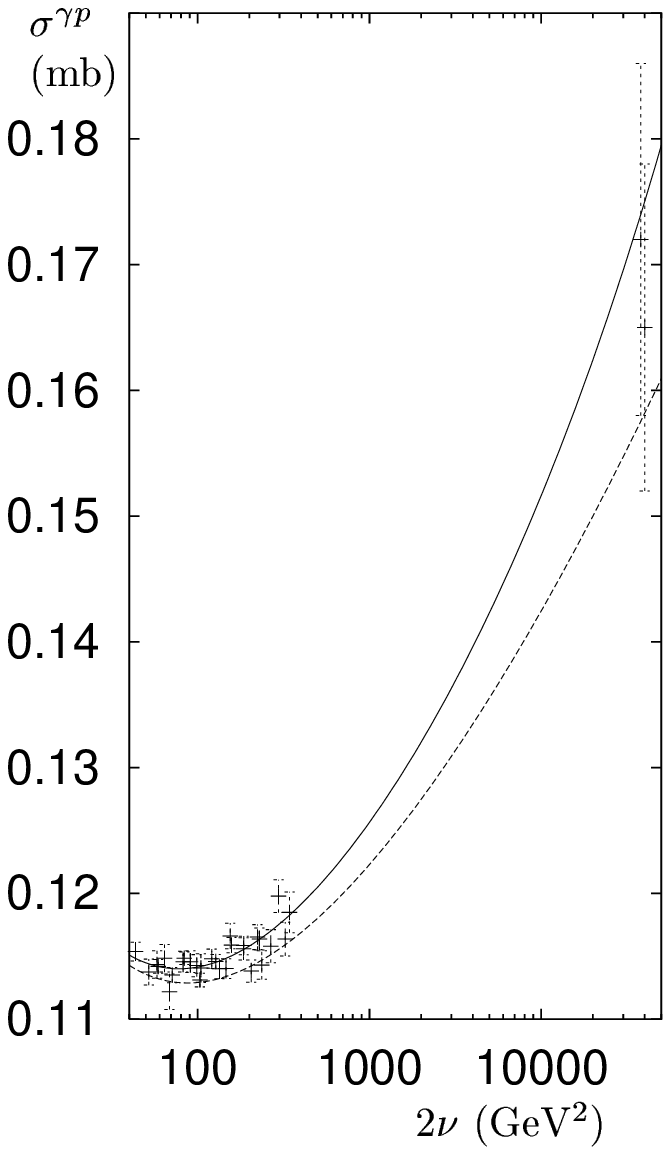}
\end{center}
\vskip -4truemm
\caption{Data for $\sigma^{\gamma p}$ with the fit obtained by
extrapolating to $Q^2=0$
the fit to $F_2(x,Q^2)$. The lower curve
has the hard-pomeron component removed.}
\label{ROUT}
\end{figure}

The total cross section for real-photon absorption is
\be
\sigma^{\gamma p}={4\pi^2\alpha\over Q^2}F_2^c(x,Q^2)\Big\arrowvert_{Q^2=0}
\ee
Figure \ref{CHARM_PHOTO} shows that hard-pomeron exchange continues
to describe charm-production data down to $Q^2=0$. 

A key question is whether all of the hard-pomeron component of
$F_2(x,Q^2)$, not just the $c$-quark part, survives at $Q^2=0$. That is,
does the hard-pomeron part of $F_2(x,Q^2)$ exist already at  $Q^2=0$,
or is it rather generated by perturbative evolution as $Q^2$ increases?
The latter view is the conventional one; I believe that it is wrong,
but the data do not yet allow us to decide one way or the other.

Certainly, it is important to include the very accurate date for
$\sigma^{\gamma p}$ in fits to $F_2(x,Q^2)$. If one does not, the
extrapolation of the fits to $Q^2=0$ misses the data by a mile.
That is, the real-photon data provide a significant constraint on
fits to $F_2(x,Q^2)$. Because these data are at comparatively low energy,
they require also an ($f_2,a_2$)-exchange term, which we included in
our fits\cite{DL01}, but its contribution to the data in figure~\ref{F2}
is extremely small. The data for $\sigma^{\gamma p}$ are shown in 
figure~\ref{ROUT}. The upper curve corresponds to  the extrapolation 
to $Q^2=0$ of the fit shown in figure~\ref{F2}, while the lower curve
shows what remains when the hard-pomeron component is omitted. 

From this figure, we cannot be sure whether a hard-pomeron component
is needed at $Q^2=0$. If the hard pomeron is present in $\sigma^{\gamma p}$
and $\sigma^{\gamma\gamma}$, then presumably it should be there at some
level also in $\sigma^{pp}$, and it will be interesting to check this
at the LHC.

\begin{figure}[t]
\vskip 3truemm
\begin{center}
\epsfxsize=0.5\textwidth\epsfbox[110 575 335 760]{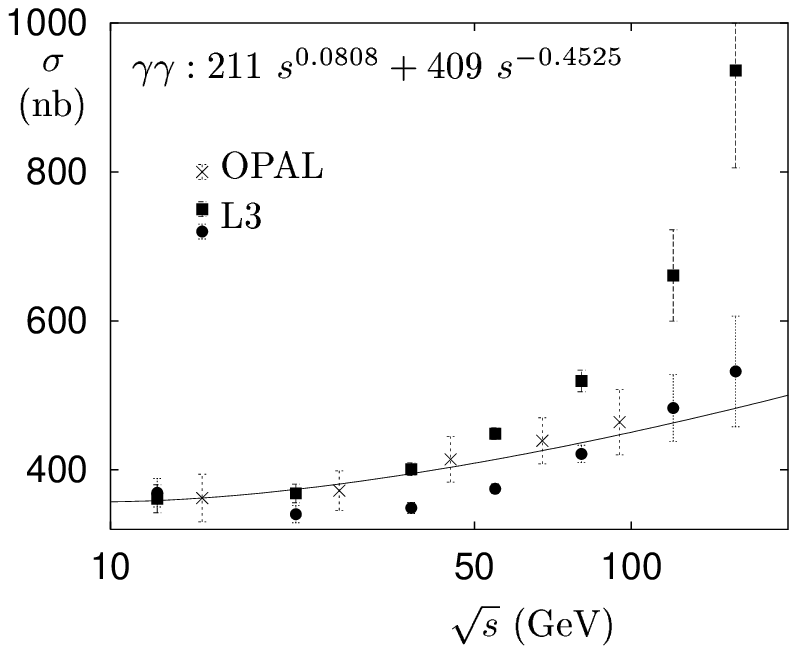}
\end{center}
\vskip -3truemm
\caption{LEP data\cite{OPAL00,L301}
for $\sigma^{\gamma \gamma}$. The curve assumes
there is no hard-pomeron contribution.}
\vskip 4truemm
\label{GG}
\end{figure}

One may ask also  whether LEP data\cite{OPAL00,L301}
for $\sigma^{\gamma\gamma}$ show evidence for
a hard-pomeron component at $Q^2=0$. Unfortunately, the data are
highly sensitive to the Monte Carlo used to correct for experimental
acceptance. Figure~\ref{GG} shows this clearly: the two sets of L3
points correspnd to different Monte Carlos. The curve is obtained
by assuming that neither $\sigma^{\gamma p}$ nor $\sigma^{pp}$
needs a  hard-pomeron term and calculating  $\sigma^{\gamma\gamma}$
by assuming that the soft-pomeron and  ($f_2,a_2$) components each
obey Regge factorisation. At present, no firm conclusion is possible!

\section{$J/\psi$ photoproduction}

\begin{figure}[t]
\begin{center}
\epsfxsize=0.55\textwidth\epsfbox[80 576 330 770]{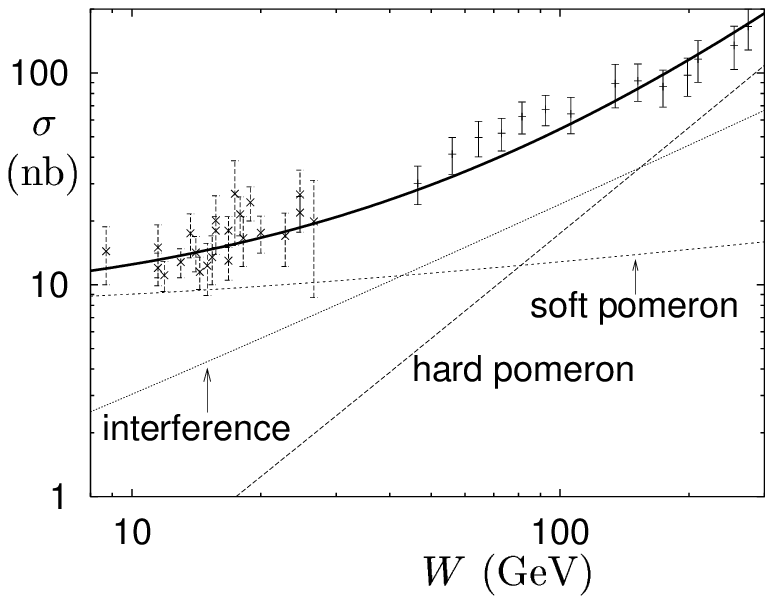}\vskip -7truemm
(a)
\vskip 9truemm
\hskip -8truemm\epsfxsize=0.65\textwidth\epsfbox[60 576 330 770]
{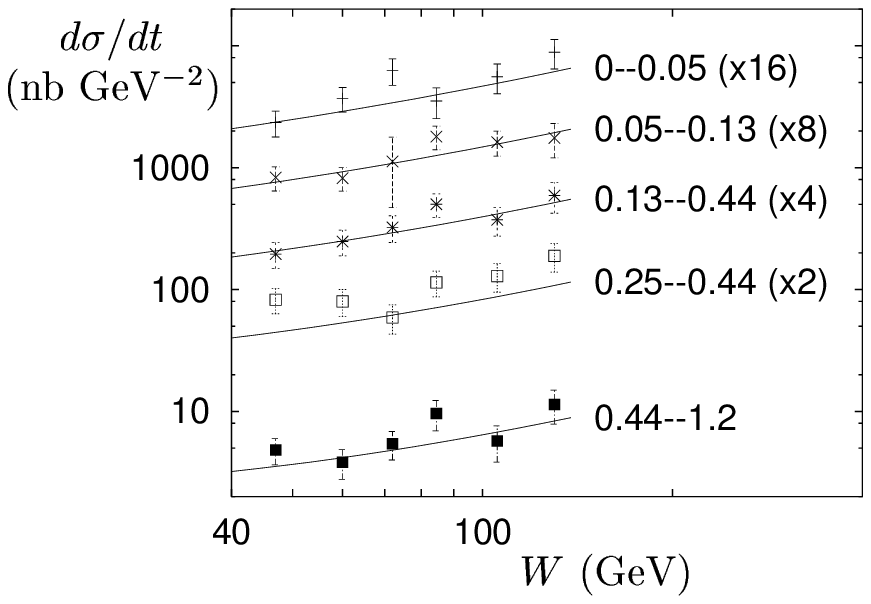}\vskip -7truemm
(b)
\end{center}
\vskip -2truemm
\caption{$\gamma p \ra J/\psi\, p$: data\ct{H100b} for
(a) the total cross section
and (b) the differential cross section.}
\label{JPSI}
\end{figure}

I have shown in figure~\ref{RHODSIGMA1}
that the rise of the cross section $\gamma p\to \rho~p$
with energy is consistent with soft-pomeron exchange, and the same is
so for $\gamma p\to \phi~p$. However, this is not true for
$\gamma p\to J/\phi~p$, for which the data\ct{H100b} require
a significant hard-pomeron component. The fits in figure~\ref{JPSI}
correspond to\cite{DL01}
\be
T(s,t) = iF_1(t)G_{J/\psi}(t)\sum_{i=0,1}
A_{\P_i}(\alpha'_{\P_i}s)^{\alpha_{\P_i}(t)-1}e^{-\half i\pi
(\alpha_{\P_i}(t)-1)}.
\label{jpsifit}
\ee
Here $F_1(t)$ is the nucleon Dirac form factor, $G_{J/\psi}(t)$ is the
coupling to the $\gamma\, J/\psi$ vertex, $\alpha_{\P_1}$ is the classical
soft-pomeron trajectory and
\be
\alpha_{\P_0}(t) = 1.44 + \alpha'_{\P_0}t
\label{e_J2}
\ee
is the hard-pomeron trajectory. We find it is sufficient to take
$G_{J/\psi}(t)$ constant 
over the range of $t$ for which data are available.
The differential cross section, figure~\ref{JPSI}b, shows little or
no shrinkage, so the slope of the hard-pomeron trajectory
$\alpha_{\P _0}(t)$ is smaller than $\alpha'_{\P _1}$, assuming that it
is linear in $t$.

It may be\cite{DL01} that in $\gamma p\to \rho~p$, $~\phi~p$ and $J/\psi~p$ the
hard-pomeron contribution is flavour-blind, and then in the first two
of these reactions it is swamped by the contribution from soft-pomeron 
exchange at $Q^2=0$, but becoming relatively more important as $Q^2$ increases.

\section{Perturbative evolution}

I have to emphasise that Regge theory is not a rival to perturbative evolution:
we have to learn how they live together. It is now accepted by most of us
that the conventional way to implement perturbative evolution at small $x$
has unsolved problems. This is because it expands the splitting matrix
${\bf P}(N,\alpha_S)$, where $N$ is the Mellin transform variable
conjugate to $x$, in powers of $\alpha_S)$, but unfortunately  the expansion
parameter turns out to be $\alpha_S/N$. Therefore the expansion becomes invalid
at small $N$ and some sort of re-summation is essential, and we do not
yet know how to do it reliably. Even though each term of the perturbative
expansion is singular at $N=0$, it is rather sure that ${\bf P}(N,\alpha_S)$
itself is not. Compare, for example, the expansion of
\be
\sqrt{N^2+\alpha_S}-N={\alpha_S/N}+O(\alpha_S^2)
\ee
which is not valid if $N<\alpha_S$.

My own guess is that ${\bf P}(N,\alpha_S)$ has no relevant singularity at
all in the complex $N$-plane. If this is true, then the DGLAP evolution 
equation
\be
Q^2{\partial {\bf u}(N,Q^2)\over \partial Q^2}=
{\bf P}(N,\alpha_S) {\bf u}(N,Q^2)
\ee
is consistent with the approximation that ${\bf u}(N,Q^2)$ just has  poles,
which are at fixed
values of $N$, so yielding fixed powers $f(Q^2)x^{-\epsilon}$ in
$F_2(x,Q^2)$ as in the fit (\ref{smallx}).
The evolution equation then tells us how $f(Q^2)$ varies
with $Q^2$ at large $Q^2$. 

\begin{figure}[t]
\begin{center}
\epsfxsize=0.4\textwidth\epsfbox[70 440 290 640]{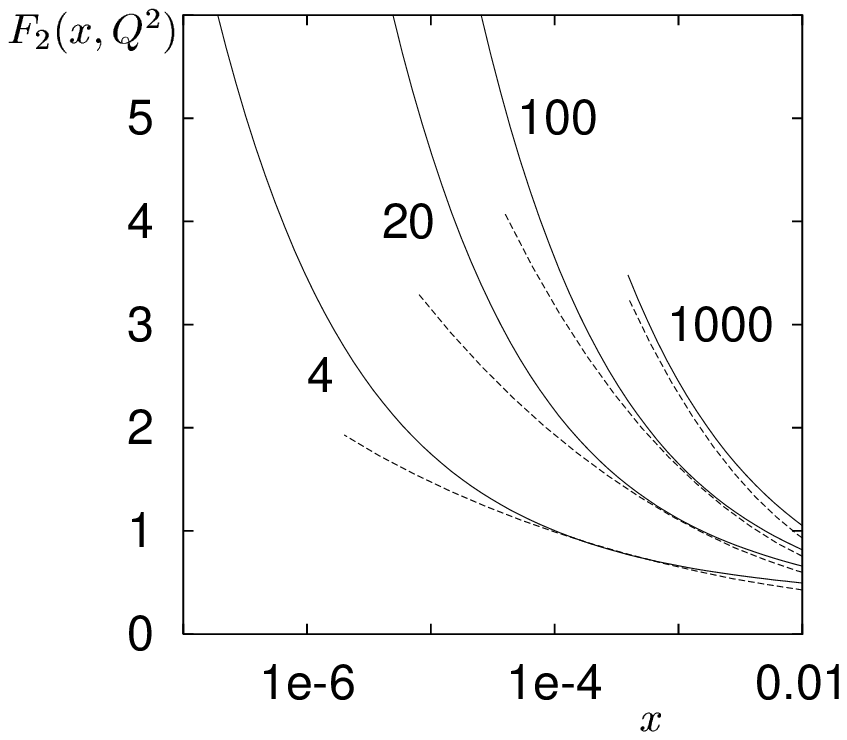}
\end{center}
\caption{Two-pomeron fit (upper curves) and two-loop-evolution fit (lower
curves)}
\vskip 4truemm
\label{EXTRAP}
\end{figure}

In the ranges of $x$ and $Q^2$ where there are data for 
$F_2(x,Q^2)$, a conventional fit based on unresummed two-loop
evolution and the two-pomeron fit can be made to agree well.
But they do not continue to agree when we extrapolate them to
smaller values of $x$, particularly at relatively small values of
$Q^2$. This is seen in figure~\ref{EXTRAP}, which compares the
two-pomeron fit (solid lines) with a two-loop-evolution fit\cite{ABF01}.

\begin{figure}[t]
\begin{center}
\epsfxsize=0.5\textwidth\epsfbox[70 550 350 760]{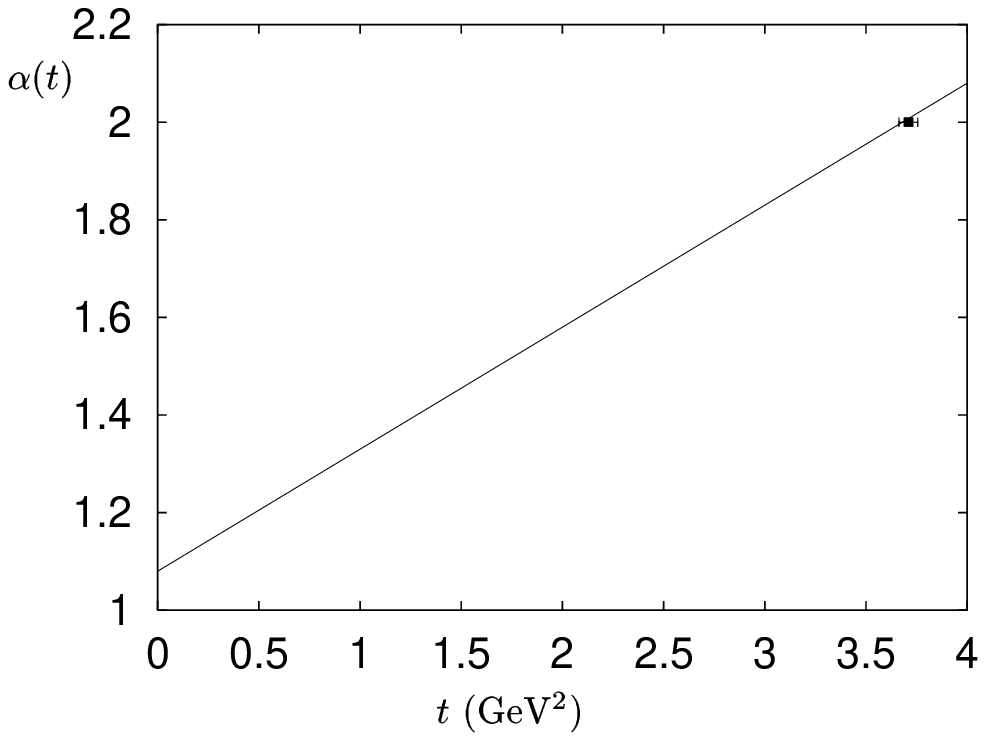}
\epsfxsize=0.42\textwidth\epsfbox[0 0 600 500]{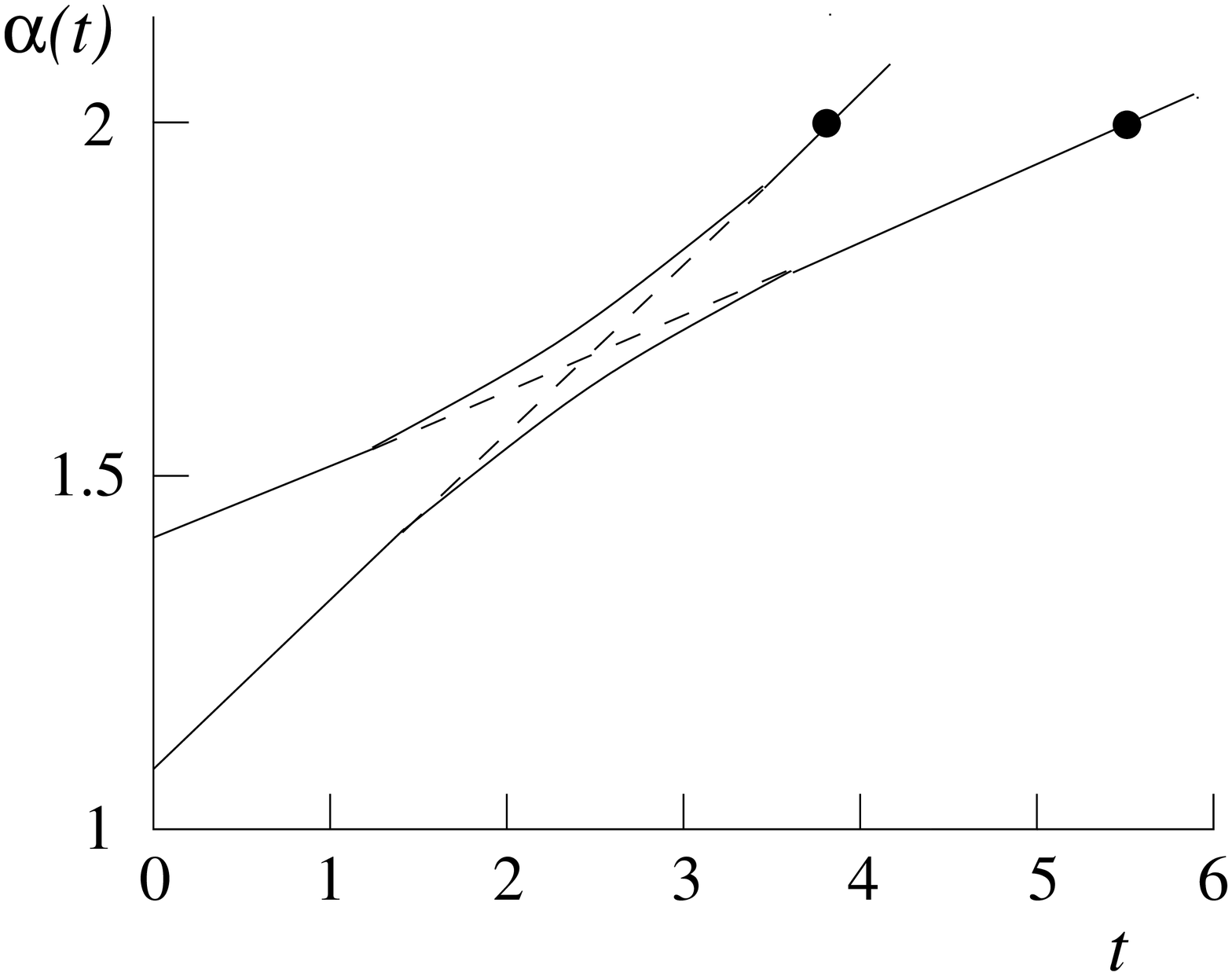}\vskip -2truemm
\end{center}
\centerline{\hfill (a)\hfill\hfill (b)\hfill}
\vskip -0truemm
\caption{Glueball trajectories:
(a) $2^{++}$ glueball candidate\ct{WA91}, with the trajectory
(\ref{softpom}); (b) repelling trajectories (\ref{softpom})
and (\ref{hardpom1}), with a second glueball candidate\cite{WA102a}}
\vskip 4truemm
\label{GLUEBALL}
\end{figure}

If indeed the hard and soft pomerons correspond to poles in the
$N$-plane, they also correspond to poles in the complex angular 
momentum plane. In this case, for the value of $t$
at which either trajectory 
passes through 2 there should be a particle with mass $\sqrt t$.
Theoretical prejudice has it that these particles should be glueballs.
Figure~\ref{GLUEBALL}a shows that there is\cite{WA91} a $2^{++}$ glueball 
candidate with exactly the right mass to be on the soft-pomeron trajectory
(\ref{softpom}).
There is a second such candidate\cite{WA102a} with the right mass
to be on the hard-pomeron trajectory (\ref{hardpom1}), but if two
trajectories cross it is likely that they mix so that they
avoid each other in the way shown in
figure~\ref{GLUEBALL}b. This picture gives the hard-pomeron trajectory a slope
close to 0.1 as in (\ref{hardpom2}).

\section{Key questions}

Among the key questions that I have raised in this talk are:

\begin{itemize}
\item Is a hard pomeron already present at $Q^2=0$?

\item Are there glueballs on pomeron trajectories?

\item Will the hard pomeron be evident in $pp$ total cross section at the LHC?

\item Can we construct the theory of perturbative evolution at small $x$?
\end{itemize}

\def\bf{}
\bibliography{blois}

\providecommand{\href}[2]{#2}\begingroup\raggedright\begin{thebibliography}{10}

\bibitem{DL92}
A~Donnachie and P.~V Landshoff,  Physics Letters {\bf B296} (1992) 227

\bibitem{CDF94}
F~Abe {\em et~al},  {\bf CDF} Collaboration, Physical Review {\bf D50} (1994)
  5550

\bibitem{CEKLT98}
J.~R Cudell {\em et~al},  Physical Review {\bf D61} (1998) 034019

\bibitem{PDG00}
{Particle Data Group},  European Physical Journal {\bf C15} (2000) 1

\bibitem{JL74}
G.~A Jaroskiewicz and P.~V Landshoff,  Physical Review {\bf D10} (1974) 170

\bibitem{Nag79}
E~Nagy {\em et~al},  Nuclear Physics {\bf B150} (1979) 221

\bibitem{Amo90}
N.~A Amos {\em et~al},  {\bf E710} Collaboration, Physics Letters {\bf B247}
  (1990) 127

\bibitem{Ast82}
D~Aston {\em et~al},  Nuclear Physics {\bf B209} (1982) 56

\bibitem{ZEUS98}
J~Breitweg {\em et~al},  {\bf ZEUS} Collaboration, European Journal of Physics
  {\bf C1} (1998) 81

\bibitem{DL98}
A~Donnachie and P.~V Landshoff,  Physics Letters {\bf B437} (1998) 408

\bibitem{DL01}
A~Donnachie and P.~V Landshoff,  hep-ph/0105088 {\bf $~$} (2001) $~$

\bibitem{ZEUS00c}
J~Breitweg {\em et~al},  {\bf ZEUS} Collaboration, Physics Letters {\bf B487}
  (2000) 53

\bibitem{H101}
C~Adloff {\em et~al},  {\bf H1} Collaboration, European Physical Journal {\bf
  C19} (2001) 269

\bibitem{H101a}
C~Adloff {\em et~al},  {\bf H1} Collaboration, European Physical Journal {\bf
  C21} (2001) 33

\bibitem{Z00}
J~Breitweg {\em et~al},  {\bf ZEUS} Collaboration, European Physical Journal
  {\bf C12} (2000) 35

\bibitem{OPAL00}
G~Abbiendi {\em et~al},  {\bf OPAL} Collaboration, European Physical Journal
  {\bf C14} (2000) 199

\bibitem{L301}
M~Acciari {\em et~al},  hep-ex/0102025 {\bf $~$} (2001) $~$

\bibitem{H100b}
C~Adloff {\em et~al},  {\bf H1} Collaboration, Physics Letters {\bf B483}
  (2000) 23

\bibitem{ABF01}
G~Altarelli, R.~D Ball and S~Forte,  Nuclear Physics {\bf B599} (2001) 383

\bibitem{WA91}
S~Abatzis {\em et~al},  {\bf WA91} Collaboration, Physics Letters {\bf B324}
  (1994) 509

\bibitem{WA102a}
D~Barberis {\em et~al},  {\bf WA102} Collaboration, Physics Letters {\bf B397}
  (1997) 339

\end{thebibliography}\endgroup
\end{document}